\documentclass{iaus}
\usepackage{graphicx}

\title[IAUS 293: Main-belt comets as tracers of ice in the inner Solar system] 
{Main-belt comets as tracers of ice \\ in the inner Solar system}

\author[Henry H.\ Hsieh]   
{Henry H.\ Hsieh$^{1,2}$}

\affiliation{$^1$Institute for Astronomy, University of Hawaii, \\ 2680 Woodlawn Drive, Honolulu, Hawaii 96822, USA \\ email: {\tt hsieh@ifa.hawaii.edu} \\
$^2$Hubble Fellow}

\pubyear{2013}
\volume{293}  
\pagerange{1--6}
\jname{Formation, Detection, \& Characterization of Extrasolar Habitable Planets}
\editors{N.\ Haghighipour, ed.}
\begin{document}

\maketitle

\begin{abstract}
As a recently recognized class of objects exhibiting apparently cometary (sublimation-driven) activity yet orbiting completely within the main asteroid belt, main-belt comets (MBCs) have revealed the existence of present-day ice in small bodies in the inner solar system and offer an opportunity to better understand the thermal and compositional history of our solar system, and by extension, those of other planetary systems as well.  Achieving these overall goals, however, will require meeting various intermediate research objectives, including discovering many more MBCs than the currently known seven objects in order to ascertain the population's true abundance and distribution, confirming that water ice sublimation is in fact the driver of activity in these objects, and improving our understanding of the physical, dynamical, and thermal evolutionary processes that have acted on this population over the age of the solar system.\looseness=-1
\end{abstract}

\firstsection 
\section{Introduction}

\begin{figure}[ht]
\begin{center}
\includegraphics[width=5.3in]{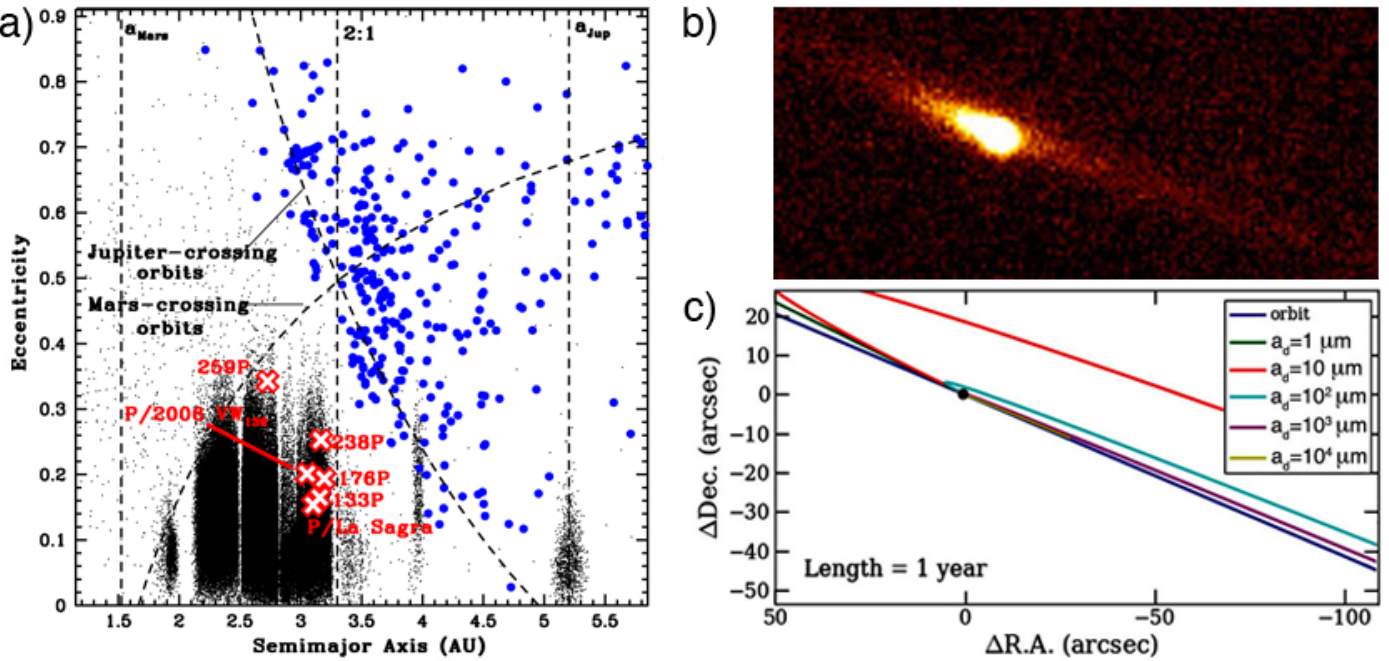} 
 \caption{(a) Semimajor axis vs. eccentricity plot of the known MBCs (red and white X symbols), all other comets (large blue circles), and asteroids (small black dots), where the MBCs can be seen to be dynamically distinct from other comets but indistinguishable from main-belt asteroids. (b) Image of MBC P/2006 VW$_{139}$ \cite[(Hsieh \etal\ 2012b)]{hsi12b}, exhibiting clearly cometary activity. (c) Syndyne plot for the P/2006 VW$_{139}$ image shown in (b) for dust grains with a range of radii ejected over the past year, showing that smaller, faster-moving dust particles would be expected to be found to the northeast of the nucleus while larger, slower-moving particles would be expected to be found to the southwest of the nucleus.
 }
\label{fig_image_orbits_fpmodel}
\end{center}
\end{figure}

Since their identification as a new cometary class \cite[(Hsieh \& Jewitt 2006)]{hsi06}, main-belt comets (MBCs) have received significant interest largely due to their potential for improving our understanding of the role of main-belt objects in the primordial delivery of volatile material (i.e., water) to the early Earth \cite[(e.g., Morbidelli \etal\ 2000, 2012; Owen 2008)]{mor00,mor12,owe08}.  As objects that exhibit apparently cometary activity (indicative of the presence of sublimating ice) yet occupy largely dynamically stable orbits in the main asteroid belt (Fig.~\ref{fig_image_orbits_fpmodel}a,b), MBCs appear to be objects that possess preserved volatile material presumably retained since the formation of the solar system.  As such, they may provide powerful constraints on the volatile content of planetesimals in the inner solar system, and in turn, on the temperature and compositional structure of the protosolar disk itself.  This information could give us insights into the conditions that allowed life to arise on Earth, and intriguingly, given recent evidence of disrupted or sublimating icy planetesimals around other stars \cite[(e.g., Lisse \etal\ 2007; Farihi \etal\ 2011; Morales \etal\ 2011)]{lis07,far11,mor11}, might even provide insights into the development of habitable conditions in extrasolar planetary systems.\looseness=-1

The first MBC, 133P/Elst-Pizarro, was discovered in 1996 \cite[(Elst \etal\ 1996)]{els96}, and since then, six more have been discovered --- 238P/Read, 176P/LINEAR, 259P/Garradd, P/2010 R2 (La Sagra), P/2006 VW$_{139}$, and P/2012 T1 (PANSTARRS) \cite[(Read \etal\ 2005; Hsieh \etal\ 2006, Garradd \etal\ 2008; Nomen \etal\ 2010; Hsieh \etal\ 2011; Wainscoat \etal\ 2012)]{rea05,hsi06,gar08,nom10,hsi11c,wai12}   --- bringing the total to seven known MBCs to date.  While most of these discoveries were made serendipitously, 176P was discovered through the targeted Hawaii Trails Project survey \cite[(HTP; Hsieh 2009)]{hsi09} which specifically set out to find MBCs, and P/2006 VW$_{139}$ and P/2012 T1 were discovered by a dedicated comet detection program within the Pan-STARRS1 wide-field survey project \cite[(Hsieh \etal\ 2012b)]{hsi12b}.  All of the currently known MBCs were discovered with telescopes smaller than 2~m in size, except for 176P which was discovered by the 8~m Gemini North telescope.  Main-belt asteroids with sufficient ice on their surfaces to drive cometary activity are unexpected, and so it has been hypothesized that MBCs may have instead preserved ice in subsurface layers over the age of the solar system, and that that ice has only recently become exposed, perhaps via excavation from an impact by another asteroid \cite[(Hsieh \etal\ 2004; Capria \etal\ 2012)]{hsi04,cap12}.\looseness=-1

The circumstances under which the currently known MBCs were discovered (i.e., using mostly small telescopes, and through mostly non-systematic search efforts) suggest that many more undiscovered active MBCs should exist.  Meanwhile, the transience of MBC activity further suggest that there should be an even larger dormant population essentially consisting of main-belt objects that do not currently exhibit observable cometary activity, yet are nonetheless icy.  In order to achieve the aforementioned insights into the formation of our solar system as well as into the circumstances that led to the rise of life on Earth and could determine the habitability of planets in extrasolar systems, we certainly need to better understand the abundance and distribution of the entire population of active and inactive MBCs.  For this information to be meaningful, however, we must first better understand the precise physical and dynamical nature of this population that we wish to use to trace ice in the inner solar system, namely (1) whether these objects are actually icy, and (2) whether they are actually representative of the inner solar system.\looseness=-1

\section{Are They Actually Icy?}


Though 133P clearly had the outward appearance of a comet, this initial discovery of an apparently icy object in the asteroid belt was surprising enough that an impact event, rather than sublimation of volatile material, was suspected of producing the object's observed comet-like activity \cite[(e.g., T\'oth 2000)]{tot00}.  However, observations of renewed activity in 2002 and 2007 \cite[(Hsieh \etal\ 2004, 2010; Lowry \& Fitzsimmons 2005)]{hsi04,hsi10,low05} rendered that explanation highly implausible.  While recurrent activity (with intervening periods of inactivity) is commonly observed for classical comets where sublimation drives activity, it would be highly improbable for impacts to occur on the same object on three separate occasions in just 11 years, each time at a similar location in the object's orbit, when impacts are not seen to occur at anywhere near that frequency or regularity on other main-belt asteroids.\looseness=-1

The ability to accurately identify the source of comet-like activity for a given object has taken on added importance with the recent discoveries of comet-like objects whose activity actually does appear to be impact-generated.  P/2010 A2 (LINEAR), (596) Scheila, and P/2012 F5 (Gibbs) were all observed to exhibit comet-like dust features which were later determined to in fact be ejecta clouds produced by recent impacts \cite[(Jewitt \etal\ 2010, 2011; Snodgrass \etal\ 2010; Bodewits \etal\ 2011; Ishiguro \etal\ 2011a,b, Stevenson \etal\ 2012)]{jew10,jew11,sno10,bod11,ish11a,ish11b,ste12}.  As such, despite the observationally comet-like morphologies of these objects, we do not consider them to be ``true'' MBCs, and instead prefer the term ``disrupted asteroids'' \cite[(e.g., Hsieh \etal\ 2012a)]{hsi12a}, where the general term ``active asteroids'' has been suggested to encompass both true MBCs and disrupted asteroids \cite[(Jewitt 2012)]{jew12}.  Making this distinction is important because when considering MBCs as tracers of the distribution of volatile material in the asteroid belt, we only wish to consider those objects for which activity is sublimation-driven, indicating the presence of ice, and thus require techniques for excluding objects whose activity is due to non-volatile-driven processes like collisions.\looseness=-1

At the present time, recurrent activity is one of the strongest (albeit indirect) pieces of observable evidence that we know of to infer whether comet-like activity is sublimation-driven or not \cite[(Hsieh \etal\ 2012a; Jewitt 2012)]{hsi12a,jew12}.  Due to the recent discovery dates of most of the known MBCs, however, recurrent activity has only been confirmed to date for 133P and 238P.
For newly-discovered MBCs for which recurrent activity has not yet been confirmed, dust modeling provides another means for deducing the source of observed activity.  For example, a syndyne plot of P/2006 VW$_{139}$ \cite[(cf.\ Finson \& Probstein 1968)]{fin68}, where each colored line represents the locus of expected positions of dust particles of a certain size ejected over the last year prior to the displayed observational data (Fig.~\ref{fig_image_orbits_fpmodel}b), is shown in Figure~\ref{fig_image_orbits_fpmodel}c.  In the image, both an anti-solar dust tail (upper left) and an orbit-aligned dust trail (lower right) can be seen, corresponding to syndynes for smaller and larger particles, respectively.  The fact that both are observed simultaneously strongly indicates that an extended emission event (consistent with sublimation, and inconsistent with impulsive impact-driven ejection) is responsible.  This is because fast-moving small particles should disperse from the nucleus within a matter of weeks, and so the fact that an anti-solar tail is still observed means that it must consist of recently ejected particles.  Large particles move slowly though, and so the length of the orbit-aligned trail indicates that those particles must have been ejected some time ago in order to have reached that distance from the nucleus by the time of the observations.  We therefore see that both old and recent dust features are present, indicating that they were likely produced by a prolonged emission process, e.g. sublimation.\looseness=-1



It should be noted that ice is not completely unexpected in main belt objects.  Theoretical studies place the snow line (the distance from the Sun beyond which temperatures in the protosolar disk were low enough for water to be able to condense into ice grains and be accreted into forming planetesimals) potentially as close to the Sun as the orbit of Mars \cite[(Sasselov \& Lecar 2000; Ciesla \& Cuzzi 2006; Lecar \etal\ 2006)]{sas00,cie06,lec06}, findings that are supported by evidence of hydrated minerals (indicative of past liquid water) in main-belt asteroids and meteorites \cite[(cf. Hiroi \etal\ 1996; Rivkin \etal\ 2002)]{hir96,riv02}.  Meanwhile, thermal modeling studies have shown that despite the fact that surface ice on main-belt objects today is highly unstable against sublimation, subsurface ice could potentially remain preserved over the age of the solar system \cite[(Schorghofer 2008; Prialnik \& Rosenberg 2009)]{sch08,pri09}.  The same impact excavation events thought to trigger current MBC activity \cite[(Hsieh \etal\ 2004)]{hsi04} will still deplete the volatile content of an asteroid's upper surface layers over Gyr timescales \cite[(Hsieh 2009)]{hsi09}, but the membership of at least two MBCs in extremely young asteroid families \cite[(Nesvorn\'y \etal\ 2008; Novakovi\'c \etal\ 2012)]{nes08,nov12} suggest that recent catastrophic disruptions of larger ice-bearing bodies could result in a significant population of objects with subsurface ice at depths shallow enough to be susceptible to impact excavation and therefore triggering of sublimation-driven activity.\looseness=-1

In terms of present-day ice, Ceres is believed to have a water content of 15\% to 30\% by mass based on shape and density data \cite[(McCord \& Sotin 2005; Thomas \etal\ 2005)]{mcc05,tho05}.  Direct spectroscopic detections of ice on the surface of asteroid (24) Themis and other large outer main-belt asteroids have been reported \cite[(Rivkin \& Emery 2010; Campins \etal\ 2010; Takir \& Emery 2012)]{riv10,cam10,tak12}, though there is some question as to whether those detections could instead be due to non-volatile minerals \cite[(e.g., Beck \etal\ 2011)]{bec11}.
\looseness=-1

\begin{figure}[ht]
\begin{center}
\includegraphics[width=5.3in]{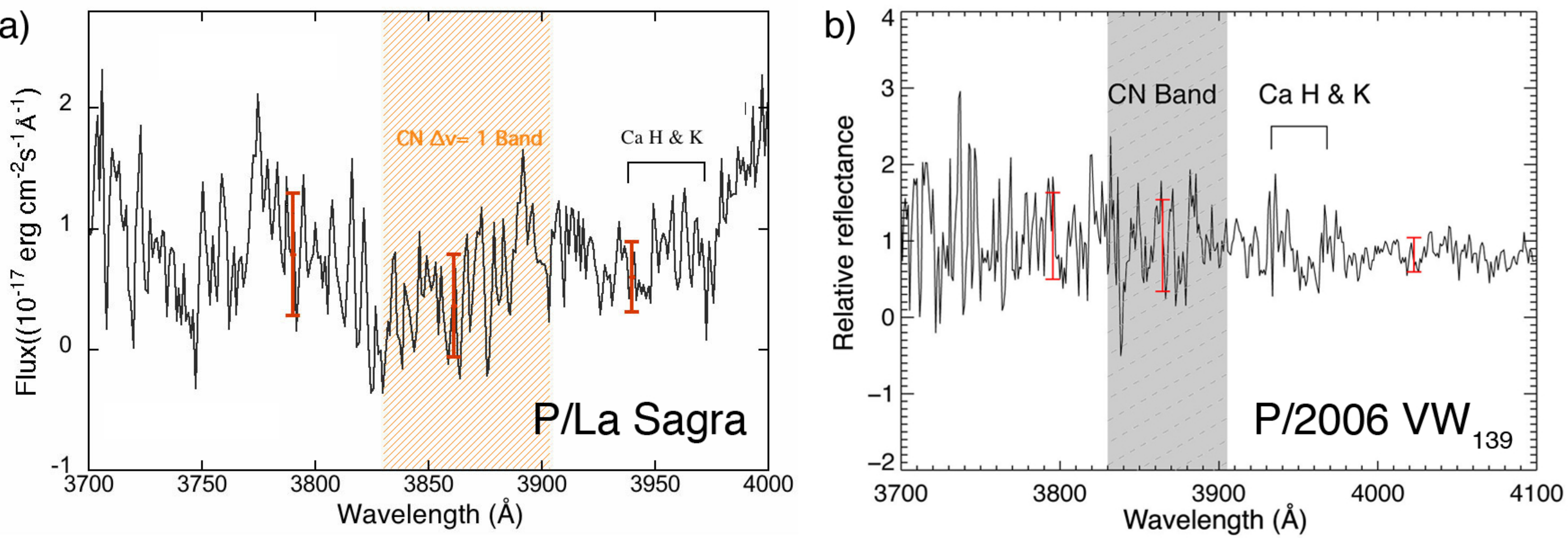} 
 \caption{Optical spectra of (a) P/La Sagra \cite[(Hsieh \etal\ 2012c)]{hsi12c}, and (b) P/2006 VW$_{139}$ \cite[(Hsieh \etal\ 2012b)]{hsi12b}, indicating the non-detection of CN emission at $\sim$3880\AA.  In each case, we found upper limit CN production rates of $Q_{\rm CN}\lesssim10^{24}$~mol~s$^{-1}$, from which we inferred approximate upper limit H$_2$O production rates of $Q_{\rm H_2O}\lesssim10^{26}$~mol~s$^{-1}$.\looseness=-1
 }
\label{fig_spectra}
\end{center}
\end{figure}

Despite the abundance of indirect evidence of water in asteroids and ice sublimation in MBCs, no direct spectroscopic detections of outgassing from a main-belt asteroid have yet been made.  Multiple attempts have been made to detect CN in MBCs using large aperture telescopes such as Keck I, the Very Large Telescope (VLT), and Gemini North and South \cite[(Jewitt \etal\ 2009; Licandro \etal\ 2011; Hsieh \etal\ 2012b, 2012c; Fig.~\ref{fig_spectra})]{jew09,hsi12b,hsi12c}, and even H$_2$O using Herschel \cite[(de Val-Borro \etal\ 2012)]{val12}, but each was only able to set upper limits on gas emission.  A search for CN emission from (24) Themis using Keck was also unsuccessful \cite[(Jewitt \& Guilbert-Lepoutre 2011)]{jew11b}.  Due to the expected extreme difficulty of detecting weak gas signatures from these relatively faint, distant objects, however, these non-detections are not considered to be definitive evidence of the lack of gas, but simply indications that the amount of any gas present was below those particular observations' detection limits.  Since even just one direct detection of gas from an MBC would be powerful confirmation that ice sublimation drives the activity of all verified MBCs, continued efforts to directly detect sublimation in MBCs using new techniques or new facilities or targeting particularly bright new targets remain extremely desirable.  Ultimately, however, detailed knowledge of MBC composition (including determination of astrobiologically significant quantities such as the D/H ratio in MBC ice) may require in situ measurements by a visiting spacecraft.
\looseness=-1

\section{Are They Actually From the Inner Solar System?}

If MBCs are to be considered reliable compositional probes of ice in the inner solar system, besides confirming that they in fact contain ice, it is clearly also essential to confirm that they are actually representative of the primordial inner solar system.  This question originally arose in the case of 133P as dynamicists attempted to determine whether this comet orbiting in the main asteroid belt could simply be an ordinary Jupiter-family comet from the outer solar system that had somehow evolved onto a main-belt orbit.  Numerical simulations showed, however, that 133P was dynamically stable over large timescales and found no plausible dynamical paths from a JFC-type orbit to that of 133P \cite[(e.g., Ipatov \& Hahn 1999; Fern\'andez \etal\ 2002)]{ipa97,fer02}.  Numerical analyses of the dynamical behavior of subsequently discovered MBCs \cite[(Haghighipour 2009; Hsieh \etal\ 2012b,c)]{hag09,hsi12b,hsi12c} produced similar results, indicating that most of the currently known MBCs are largely stable, and therefore likely diagnostic of the local composition of the protosolar disk.  MBC 259P was found to be unstable \cite[(Jewitt \etal\ 2009)]{jew09}, however, indicating that it is likely {\it not} indicative of the composition of the protosolar disk at that location, underscoring the need to evaluate the dynamical stability of each new MBC as more are discovered.\looseness=-1

Despite their dynamical stability in the present-day solar system, the suitability of MBCs as compositional probes of the inner protosolar disk has been challenged by recent large-scale dynamical modeling results suggesting that outer solar system material could have been delivered to the asteroid belt early on as the solar system was still evolving towards its current dynamical configuration.  \cite[Levison \etal\ (2009)]{lev09} noted that one of the consequences of the so-called Nice model of the formation of the solar system is that perturbations from giant planet migration could have resulted in significant contamination of the asteroid belt by icy trans-Neptunian objects (TNOs), raising the possibility that the likewise icy MBCs could be from the outer solar system after all.  Levison \etal\ expect that TNO interlopers would remain taxonomically distinct though from ``native'' main-belt material, appearing as P- and D-type asteroids (much like the currently known Hilda and Jovian Trojan asteroids).  However, the MBCs which have been taxonomically classified to date are identified as B-type asteroids, a sub-class of the C-type class of asteroids considered to be ``typical'' in the outer asteroid belt \cite[(Licandro \etal\ 2011)]{lic11}.\looseness=-1

Somewhat more problematically, the recently proposed ``Grand Tack'' model of early solar system evolution, in which the giant planets first migrate inward and then outward, suggests that C-type asteroids could also originate in the outer solar system (outside the orbit of Jupiter) \cite[(Walsh \etal\ 2011)]{wal11}.  The model however suggests that the same outer solar system C-type material that was implanted in the outer asteroid belt during the ``Grand Tack'' period was also placed onto terrestrial planet-crossing orbits, meaning that icy objects in the outer asteroid belt (i.e., MBCs) might still have played a key role in the primordial delivery of terrestrial water.  The possibility that these objects may not be native to the asteroid belt, however, means that care must be taken when trying to use them to apply constraints on protosolar disk models.  Clearly, more theoretical and (ideally) observational work to confirm or refute the Grand Tack model (and other future models of its type) will be necessary to ensure the proper interpretation of any compositional and temperature information gleaned from MBCs as their overall spatial distribution becomes better determined in the future as more objects are discovered.

\section{Looking Ahead}

At the current time, significant work remains before MBCs can be used to accurately trace the distribution ice in the inner Solar system.  In addition to the issues raised above, the most critical current deficiency is that with only seven objects known to date, we lack a complete picture of the total abundance and distribution of the MBCs themselves.  Previous untargeted searches have been attempted but these have suffered from limited data and a lack of real-time follow-up capability \cite[(Gilbert \& Wiegert 2009, 2010; Sonnett \etal\ 2011)]{gil09,son11}. While these surveys were untargeted, it is also important to note that they were not unbiased, as their samples were inevitably skewed towards larger and closer (e.g., inner main belt) asteroids due to observational selection effects.  New high-sensitivity, large-scale surveys already in operation (e.g., PS1) or currently in development (e.g., the Large Synoptic Survey Telescope, or LSST) hold greater promise for detecting more MBCs but effective and efficient comet detection algorithms, able to operate in real time on millions of detections each night, coupled with complementary observing programs capable of providing timely follow-up, will be necessary to realize the full discovery potential of these surveys.  Targeted observations of objects considered to be high-probability MBC candidates \cite[(e.g., Hsieh 2009)]{hsi09} and searches for MBCs in archival data may also prove to be useful supplements to the search efforts of large surveys.\looseness=-1

Related to the problem of ascertaining the true extent of the population of active MBCs is ascertaining the extent of the population of {\it inactive} MBCs, i.e., ice-bearing asteroids that do not currently exhibit observable cometary activity.  Assuming that MBC activity is in fact produced by the sublimation of localized impact-excavated pockets of exposed subsurface ice and that the volatile lifetimes of these exposed pockets are finite, alongside the active MBC population, there must be a related and certainly much larger population of asteroids containing subsurface ice but that simply have not experienced recent excavating impacts that would produce observable evidence of that ice.  Determination of the extent of this inactive population will likely rely on developing an improved understanding of the nucleus properties of currently known MBCs, e.g., through observations at aphelion when most MBC activity is seen to largely subside.  A reliable means for identifying icy asteroids in the absence of activity (which is transient even in the case of active MBCs) would enable us to use either the aforementioned untargeted wide-field surveys such as PS1 or LSST, or perhaps smaller-scale, intelligently targeted surveys to ascertain the distribution of icy asteroids, both active and inactive, in the asteroid belt.\looseness=-1

In the meantime, continued work is needed on the aforementioned fundamental issues of verifying that MBC activity is in fact correlated with the presence of ice, and determining the dynamical history of the MBCs (both as individual objects and as a population) to understand how the primordial spatial distribution of icy planetesimals may have evolved to produce the present-day distribution.  These objects have not simply undergone dynamical evolution though, of course, and as such, other processes that have acted on MBCs over time, such as thermal depletion of volatiles, collisional depletion, and catastrophic disruptions \cite[(cf.\ Schorghofer 2008; Hsieh 2009; Nesvorn\'y \etal\ 2008; Novakovi\'c \etal\ 2012)]{sch08,hsi09,nes08,nov12}, will also need to be studied and well-understood in order for useful information to be inferred about the primordial protosolar disk from the MBCs.

\end{document}